\documentclass{aa}
\usepackage{graphicx}

\begin{document}
\title{Obscured clusters.\,IV. The most massive stars in [DBS2003]\,179. \thanks{Based on observations gathered  with ISAAC,VLT and VISTA of the ESO and the 8.2-meter GEMINI telescope within observing programs 81.D-0471, 179.B-2002, and GS-2009A-Q-81.}}
\author{J. Borissova\inst{1,2} 
   \and
   L. Georgiev\inst{3}
   \and
   M. M. Hanson \inst{4}
   \and
   J. R. A. Clarke \inst{1}
   \and
   R. Kurtev\inst{1}
   \and
   V.D. Ivanov\inst{5}
   \and
   F.Penaloza\inst{1,2}
   \and
   D. J. Hillier\inst{6}
   \and
   J. Zsarg\'o\inst{7}
}
    
\offprints{J. Borissova}

\institute{Departamento de Fis\'ica y Astronom\'ia, Facultad de
  Ciencias, Universidad de Valpara\'iso, Av. Gran Breta\~na 1111,
  Playa Ancha, Casilla 5030, Valpara\'iso, Chile \\
    \email{jura.borissova@uv.cl; radostin.kurtev@uv.cl}
\and
{The Milky Way Millennium Nucleus, Av. Vicu\~{n}a Mackenna 4860, 782-0436 Macul, Santiago, Chile}  
\and
  Instituto de Astronomia, Universidad Nacional Aut\'onoma de M\'exico,
  Apartado Postal 70-254, CD Universitaria, CP 04510 Mexico DF, Mexico,  \\
    \email{georgiev@astro.unam.mx} 
\and
Department of Physics,
University of Cincinnati, Cincinnati, OH 45221-0011, USA \\ 
    \email{margaret.hanson@uc.edu}
\and
  European Southern Observatory, Ave. Alonso de Cordova 3107,
  Casilla 19, Santiago 19001, Chile \\
    \email{vivanov@eso.org}
\and
Department of Physics and Astronomy, University of Pittsburgh, Pittsburgh, PA, 15260 USA\\
\email{Hillier@pitt.edu}
\and
Escuela Superios de Fisica y Matematicas, Instituto Politecnico Nacional, Mexico \\
\email{jzsargo@esfm.ipn.mx}
}

\date{Received; accepted}

\abstract
{}
{We report new results for the massive evolved and main sequence members of the young galactic cluster DBS2003\,179. We determine the physical parameters and investigate the high-mass stellar content of the cluster, as well as of its close vicinity.}
{Our analysis is based on ISAAC/VLT moderate-resolution (R$\approx$4000) infrared spectroscopy of the brightest cluster members. We derive stellar parameters for sixteen of the stellar members, using full non-LTE modeling of the obtained spectra.}
{ The cluster contains three late WN or WR/LBV stars (Obj\,4, Obj\,15, and Obj\,20:MDM\,32) and at least 5 OIf and 5 OV stars. According to the Hertzsprung-Russell diagram for DBS2003\,179, the WR stars show masses above 85\,$\cal M_\odot$, the OIf stars are between 40 and 80\,$\cal M_\odot$, and the main sequence O stars are $>$20\,$\cal M_\odot$. There are indications of binarity for Obj\,4 and Obj\,11, and Obj\,3 shows a variable spectrum.  The cluster is surrounded by a continuous protostar formation region most probably triggered by DBS2003\,179.}
{We confirm that DBS2003\,179 is young massive cluster (2.5\,$10^4$\,$\cal M_\odot$) very close to the Galactic center at the distance of 7.9$\pm0.8$ kpc.}


\authorrunning{J. \,Borissova et al.}
\titlerunning{[DBS2003]\,179}

\maketitle
%

\section{Introduction}

Where and how are massive stars forming in the Galaxy?   Until recently, this question had not instigated  significant investigation.  However, the past decade has seen a surge in studies using recently available infrared surveys to search for deeply embedded massive stars and massive clusters. In 2001, at the time of the first data release of the 2MASS survey (Skrutskie et al. 2006), numerous new cluster candidates were being discovered (Dutra \& Bica \ 2001; Dutra et al.\ 2003; Froebrich et al.\ 2007), but no one had expected to find very high mass young clusters. A few seemingly peculiar, but clearly very massive, young clusters had been identified near the center of the Galaxy (e.g. Cotera et al.\ 1996, Figer et al.\ 1997, Figer et al.\ 1999), but how they connected to {\sl normal} stellar clusters forming in the disk of the Milky Way was not yet known.   Moreover, while efficient for studying 1-2 kpc from the Sun (Carpenter 2000), the 2MASS survey did not probe the inner region, where star formation is at its greatest, and would not detect massive clusters if they were there (Hanson et al.\ 2010).  Nevertheless, astronomers have discovered very massive clusters, rivaling anything known in the Local Group of galaxies in our Milky Way.  Several new infrared surveys, such as GLIMPSE (Benjamin 2003) and VVV (Minniti et al. \ 2010, Saito et al.\ 2012), are providing deeper imaging and greater spatial resolution, which has unleashed a venerable ``cottage industry'' of independent groups working to determine the massive cluster content of our Galaxy (Homeier et al.\ 2003, Bik et al.\ 2005, Mercer et al.\ 2005, Froebrich et al.\ 2007, Davies et al.\ 2007, Messineo et al.\ 2009, Mauerhan, Van Dyk \& Morris \ 2011, Borissova et al.\ 2011).  

These all show that the Milky Way has many high-mass, young clusters, as predicted by Hanson (2003) based on the massive cluster populations found in external galaxies with Milky Way properties. But why is finding, cataloging and determining the properties of these massive clusters so important?   These young massive clusters are critical to our understanding of massive star evolution.  The upper region of the Hertzsprung-Russell diagram contains different types of emission-line stars, including Of supergiants, luminous blue variables (LBVs), O3If$/$WN6 and Ofpe$/$WN9 stars (also called hot and cool slash stars), B[e], and Wolf-Rayet (WR) stars. All these classes represent different phases in the evolution of very massive stars. However, although it is commonly accepted that very massive O stars ultimately evolve into WR stars, the details of the evolutionary connections in the intermediate stages are still poorly understood. Their formation and early evolution are hard to investigate, mainly because this period of the star's life is very short. The most direct way to investigate them is to identify and study massive stars and especially WR stars within star clusters, since this enables us to study a coeval population with uniform metallicity. For this reason, the most massive members of young galactic clusters deserve special attention.
    
The open cluster candidate [DBS2003]\,179 (hereafter DBS2003\,179) was discovered by Dutra et al. (2003) via visual inspection of 2\,MASS images during their survey of embedded clusters in the areas of known radio and optical nebulae. The cluster is located in the direction toward the H{\sc II} ionized region G347.6+0.2. Borissova et al. (2005; hereafter Paper~I) obtained high-quality infrared (IR) photometry and confirmed that DBS2003\,179 is a member of the large family of ``newly'' discovered heavily obscured clusters. Based on the photometric analysis and on the so-called ``10$^{\rm th}$ brightest star method'' (Dutra et al 2003), we determined some of the cluster parameters: $A_V$=19, ($m$$-$$M$)$_0$=13.5\,mag, age of $\sim$7\,Myr, and total mass of $\sim$5.5\,10$^3$\,$\cal M_{\odot}$. It was the first indication that the cluster is young and  massive. However, the photometry alone is not enough to derive an accurate distance to the star cluster, and the uncertainty of the tenth brightest star method can easily be as much as $\sigma${($m$$-$$M$)$_0$}$\sim$2.5\,mag or a factor of $\sim$10 in distance (see the discussion in Paper~I). To determine more accurate physical parameters for the cluster, we obtained medium-resolution ($R$=9000) $K$-band spectra with the Infrared Spectrometer And Array Camera (ISAAC) (Moorwood et al. 1998) on the ESO Very Large Telescope (VLT) at Paranal, Chile, of selected cluster members and reported seven O stars and an Ofpe/WN9 star (Borissova et al. 2008; hereafter Paper~II). Based on this we have calculated a distance of D$\sim$7.9\,kpc and a cluster mass approaching 10$^4$\,$\cal M_\odot$. The age of the cluster was estimated to be between 2 and 5\,Myr. Mauerhan, Van Dyk \& Morris (2011) reported another WN8-9 star associated with the cluster (the identification number MDM32) as part of their search for new Galactic WRs using the method of infrared color selection, and cross-correlation with X-ray point source catalogs. 

They calculated a distance to DBS2003\,179 of 6.4\,kpc, using the calibration photometry values for WN8-9 stars given in Crowther et al. (2006). In contrast, in the most recent catalog of Milky Way high-mass clusters,Davies et al. (2012), report a kinematic distance of 9.0\,kpc to the DBS2003\,179. Thus according to Davies et al. (2012) the cluster holds a rather unique location on the far side of our Milky Way and possibly aligned with the Galactic Bar.  

A more in-depth investigation, using new and improved infrared spectroscopy of key cluster members, would more firmly determine its mass, but also its distance and its possible association with the Galactic Bar. In this paper we report some new results for the massive main sequence and evolved cluster members. 
 
\section{Sample selection}

\begin{figure}
\resizebox{9cm}{!}{\includegraphics{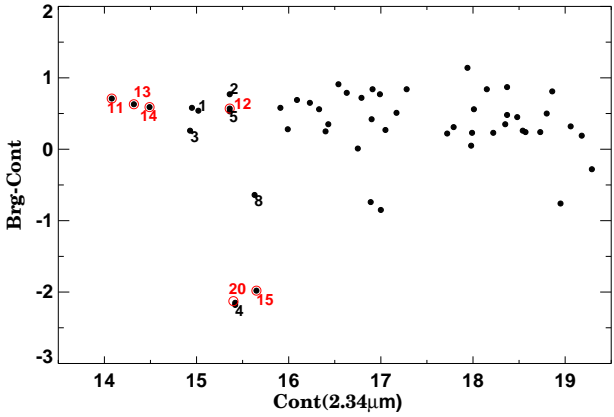}}
\caption{Results from the narrow-band photometry of
DBS2003\,179 at Br$\gamma$ and Cont(2.34$\mu$m). The labels
indicate the stars with known spectral types from Paper II. The emission line star candidates 
are marked with red circles and are labeled.
}
\label{db179_brg}
\end{figure}

In Paper II we used medium-resolution $K$-band spectra of DBS2003\,179 obtained with the ISAAC on VLT at ESO to classify seven stars showing emission lines. Four of them (Obj\,1, Obj\,2, Obj\,3 and Obj\,5) are classified as early O4-O6 V type stars, Obj\,6 is most probably a young stellar object (YSO), while Obj\,8 and Obj\,4 are O4If-O5If and Ofpe/WN9, respectively. Since the cluster seems massive, we searched for additional emission-line stars.  We constructed the narrow-band Br$\gamma$ vs. 2.34\,$\mu$m diagnostic color-magnitude diagram shown in Fig.~\ref{db179_brg}. The Br$\gamma$ images were taken with Persson's Auxiliary Nasmyth Infrared Camera (PANIC) at the 6.5-meter Baade telescope at the Las Campanas Observatory, while for the continuum we used a 2.34$\mu$m ISAAC acquisition image (for more details on the photometric observations and data reduction see Papers I and II). All magnitudes are in the instrumental systems. The stars with the spectral classification from Paper II are labeled. The diagnostic diagram indeed shows large Br$\gamma$-Cont excess for Obj\,4 and Obj\,8, the stars with spectroscopically confirmed Br$\gamma$ emission. The strong Br$\gamma$ excess of Obj\,8 is puzzling, because the object does not show such strong spectral emission. Obj\,6 is too faint and is not visible in the continuum image. Using the color excess of Obj\,8 as a cut-off limit, as well as the position of the known emission line stars on the diagnostic diagram, we selected six candidate Br$\gamma$ emission line stars. They are given in Table~\ref{candidates} and are labeled as Obj\,11, Obj\,12, Obj\,13, Obj\,14, Obj\,15, and Obj\,20. Three  additional stars, which fell into the long slit (120\,arcsec length) during the observations, namely Obj\, 17, \,18, and \,19, are also given (see Sect.~4). As pointed out, the Obj\,20 has recently been reported by Mauerhan, Van Dyk \& Morris (2011).

\begin{table*}\small
\begin{center}
\caption{Positional and photometric data for spectroscopically observed stars in DBS2003\,179.}
\label{candidates}
\begin{tabular}{lllllllll}
\hline
\multicolumn{1}{l}{Object} &
\multicolumn{1}{l}{RA(J2000)} &
\multicolumn{1}{l}{DEC(J2000)}&
\multicolumn{1}{c}{$J$} &
\multicolumn{1}{c}{Err\,$J$} &
\multicolumn{1}{c}{$H$} &
\multicolumn{1}{c}{Err\,$H$} &
\multicolumn{1}{c}{$K_{\rm S}$} &
\multicolumn{1}{c}{Err\,$K_{\rm S}$} \\
       &hh:mm:ss&deg:mm:ss& mag &  & mag & &mag \\
\hline
 Obj\,4 &	17:11:31.88& $-$39:10:46.9&		12.08&	0.01&	10.14&	0.03&	9.18&	  0.02\\	 
 Obj\,11&	17:11:31.33& $-$39:10:55.3&		13.40&	0.07&	11.46&	0.06&	10.47&	0.05\\	 
 Obj\,12&	17:11:31.55& $-$39:10:53.0&		14.49&	0.05&	12.53&	0.02&	11.63&	0.04\\	 
 Obj\,13&	17:11:31.71& $-$39:10:51.1&		13.69&	0.02&	11.65&	0.02&	10.72&	0.02\\	 
 Obj\,14&	17:11:31.67& $-$39:10:46.9&		13.72&	0.02&	11.73&	0.03&	10.76&	0.03\\	 
 Obj\,15&	17:11:31.80& $-$39:10:46.8&		12.50&	0.02&	10.53&	0.03&	9.57&	  0.03\\	 
 Obj\,17&	17:11:31.69& $-$39:10:44.4&		14.90&	0.03&	12.85&	0.02&	11.93&	0.03\\	 
 Obj\,18&	17:11:31.85& $-$39:10:29.6&		15.46&	0.03&	14.00&	0.05&	13.18&	0.03\\	 
 Obj\,19&	17:11:32.01& $-$39:10:20.1&		14.69&	0.06&	13.02&	0.04&	12.20&	0.05\\	 
 Obj\,20(MDM32)&	17:11:33.03& $-$39:10:39.7&		12.37&	0.02&	10.43&	0.03&	9.19&	  0.02\\
\hline
\end{tabular}
\end{center}
\end{table*}

\begin{figure}[h]
\includegraphics[width=\columnwidth]{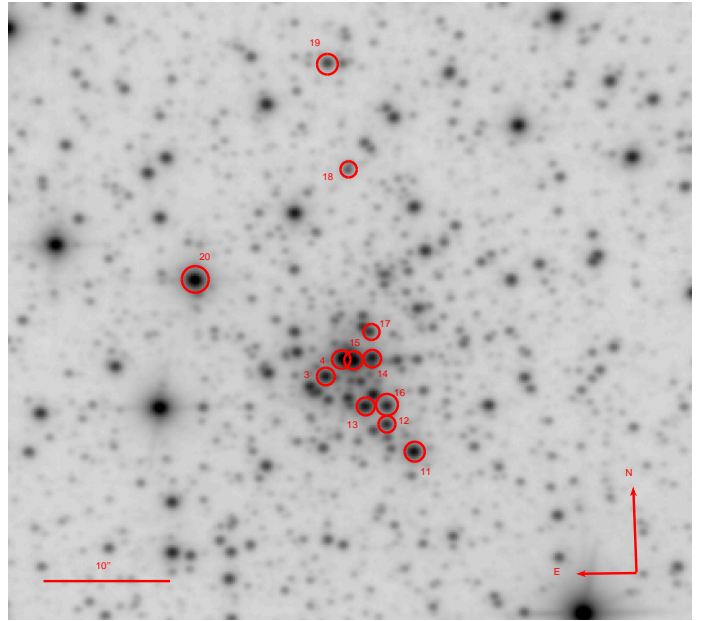}
\caption{Identification of the stars with near-IR spectroscopy listed in Table~\ref{candidates}. This is an ISAAC $K_{\rm S}$ band image with field of view of approximately 1$\times$1\,arcmin. North is up, east to the left.
}
\label{db179_chart}
\end{figure}

\section{Observations and data reduction}\label{obs_data_red}

Medium-resolution spectra of the DBS2003\,179 were obtained at the 8.2-m Unit 1 telescope of the ESO VLT, located on Cerro Paranal in Atacama, Chile, with ISAAC (Moorwood et al. 1998). ISAAC employs a Rockwell Hawaii 1024$\times$1024 array.  The pixel scale of the detector is 0.146 arcsec pixel$^{-1}$. Our spectra were obtained with the slit width of 0\farcs6, giving a spectral resolution of approximately 4000 (slightly more in the $K$-band, and slightly less in the $H$-band). The grating was centered on two positions, 1.705 and 2.150 $\mu$m in the $H$ and $K$-bands respectively. 
Unfortunately, some technical problems made it impossible to obtain the  $H$-band spectra for one of the slit positions (containing the stars Objs\,11, \,12, \,13, and \,20).  The  broadband $JHK_{\rm S}$ imaging of DBS2003\,179 was also carried out with the Aladdin detector of ISAAC during the same nights. 

The near-infrared, high-resolution ($R$$\approx$50000) spectra were obtained over five nights in February-June, 2009 with the Phoenix echelle spectrometer at Gemini-South observatory with service mode in the $K$-band spectral range. Phoenix uses an Aladdin 512$\times$1024 InSb array as the detector, and it is operable throughout the range of sensitivity of InSb, 1--5 microns. The grating is a 63.4 degree echelle with 32 lines per millimeter giving a dispersion of about 2\,10$^5$ per pixel. Since Phoenix covers a very narrow wavelength range (0.5$\%$ of the central wavelength), we chose three spectral settings centered on  2.104, 2.143, and 2.179\,$\mu$m.

For the spectroscopic reduction, we used a combination of the Image Reduction and Analysis Facility (IRAF)\footnote{IRAF is distributed by the National Optical Astronomy Observatories, which are operated by the Association of Universities for Research in Astronomy, Inc., under cooperative agreement with the National Science Foundation}, ESO {\tt eclipse} package (Devillard 2001), and the Gemini pipeline. The basic data reduction included the flat-fielding, sky subtraction, extraction of one-dimensional spectra, wavelength calibration, and telluric correction (for more details see Hanson et al. 2010). The images were reduced following the typical procedures for the infrared, which included flat-fielding, sky subtraction, alignment, and combination into a final image. The photometry was obtained using the {\tt DAOPHOT} package within IRAF. The photometric calibration was performed by comparing our instrumental magnitudes with the 2MASS measurements. The errors were calculated, taking the photometric errors and errors from the transformation to the standard system into account. These magnitudes and errors obtained for the spectroscopic targets are given in Table~\ref{candidates}.

\section{Analysis of the spectroscopic targets.}

\subsection{Spectral classification.}

\begin{figure}[h]
\resizebox{\hsize}{!}{\includegraphics{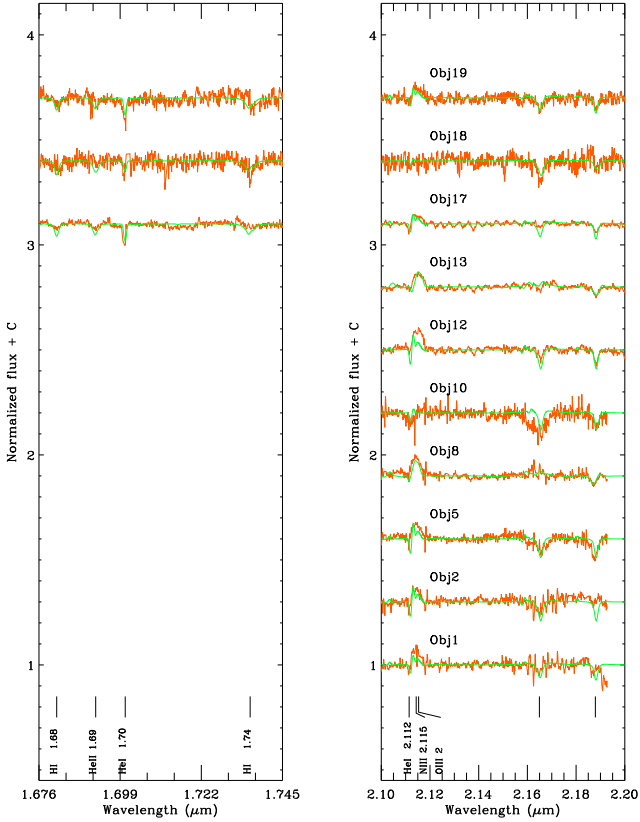}}
\caption{The ISAAC spectra of DBS2003\,179 stars. The different spectra have been arbitrarily shifted along the flux $F_{\lambda}$ axis for clarity. The dashed lines indicate the rest-frame wavelengths of the spectral features. The best-fitting models (green) are overplotted on the observed spectra (red).}
\label{db179_abs}
\end{figure}

\begin{figure}[h]
\resizebox{\hsize}{!}{\includegraphics{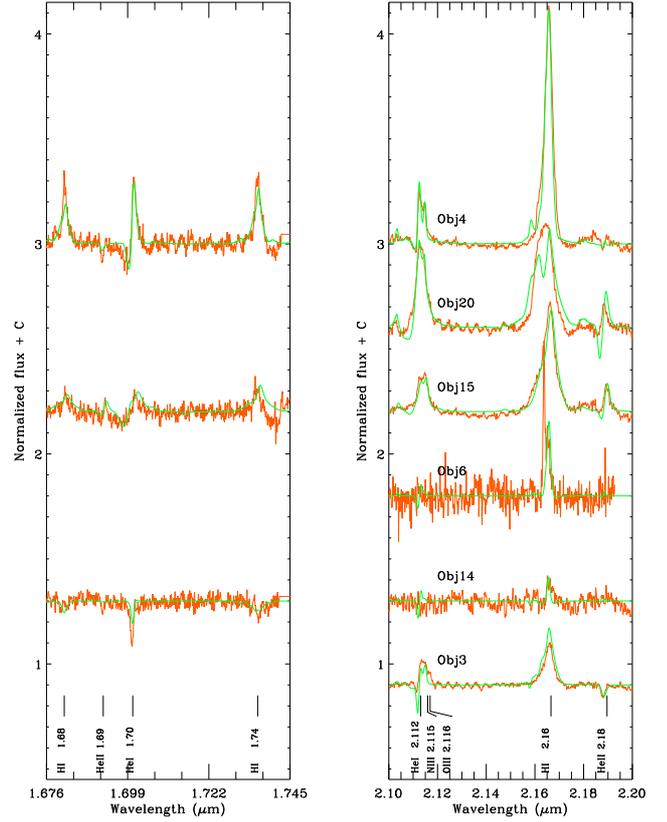}}
\caption{The same as for Fig.~\ref{db179_abs} but for  the stars with strong emission lines.}
\label{db179_emi}
\end{figure}

The spectra of candidate emission line stars given in Table~\ref{candidates} are plotted in Figs.~\ref{db179_abs} and \ref{db179_emi} in $K$ and $H$ (where available) bands.  The preliminary raw spectral classification was done using available catalogs of $K$-band spectra of objects with spectral types derived from optical studies (Morris et al. 1996; Hanson et al. 1996; Figer et al. 1997; Hanson et al. 2005), as well as from the spectral catalogs of Martins \& Coelho (2007), Crowther et al. (2006), Liermann, Hamann \& Oskinova (2009), Mauerhan, Van Dyk \& Morris (2011), and Davies et al. (2011). The most prominent lines in the $K$ and $H$-band region (Brackett\,$\gamma$ (4-7) 2.1661\,$\mu$m (Br$\gamma$); He{\sc ii} $\lambda$2.188 (7-10); He{\sc i} $\lambda$2.1127 ($3p\ ^3$P$^o-4s\ ^3$S, triplet); He{\sc ii} $\lambda$1.692 (12-7); He{\sc i} $\lambda$1.702 ($1s^4d^3D^e-1s^3p^3P^o$); H{\sc i} $\lambda$1.668 (11-4) ; H{\sc i} $\lambda$1.722 (10-4)) were compared with the template spectra from these papers.

The equivalent widths (EWs) of DBS2003\,179 spectroscopic targets are listed in Table~\ref{stars_spectra}. They were measured on the normalized continuum spectra, by the IRAF task {\tt splot} using the deblending function. The errors of the EW were estimated considering the signal-to-noise ratio measured at 2.134 $\mu$m  (see Col. 2 of Table~\ref{stars_spectra}), the peak over continuum ratio of the line (see Bik et al. 2005) and the error from the telluric star subtraction (which was estimated to be $\sim$10-15\% in the worst cases). The EW of the emission lines are negative, and the symbol ``NP'' means that the line is not present in the spectrum.

We derived the radial velocities of all stars with the
IRAF task {\tt fxcor}, which uses a cross-correlation Fourier
method. Since the stars show very different characteristics,
several radial velocity standards were used: for objects 1, 2, 3, 5, 12, 13, 17, and 19 we used 
HD15629, an OV5 spectral type star with V$_{\rm rad}$=$-$48\,km\,s$^{-1}$; Obj\,8 was
cross-correlated with HD15570, an O4If spectral type star with
V$_{\rm rad}$=$-$15\,km\,s$^{-1}$; Obj\,6 and Obj\,14 were measured
by comparison with DM+493718, a Be spectral type star; all absorption 
line stars were compared with HD191639 (B1V, V$_{\rm rad}$=$-$7\,km\,s$^{-1}$)
and HD167785 (B2V, V$_{\rm rad}$=$-$10.6\,km\,s$^{-1}$). The
spectra were taken from Hanson et al (1996, 2005) spectral atlases. It is well known that the WR stars 
show broad and blended lines, which complicate the calculations of accurate radial velocities, 
so to estimate their radial velocities we used the IRAF task {\tt rvidlines}.
The task measures radial velocities from spectra by determining the wavelength shift in 
spectral lines relative to specified rest wavelengths. The basic usage consists of identifying one or more spectral lines, entering the rest wavelengths, and computing the average wavelength shift converted to a radial velocity. Thus the user can select the lines, which means that for WR stars it is possible to reject the broad and emission lines, which are not suitable for RV measurements. 
The lines are centered using Gaussian fits and when available, the purest line in our K band spectra, He{\sc ii} $\lambda$2.188, is taken as reference.

\begin{figure}
\resizebox{\hsize}{!}{\includegraphics{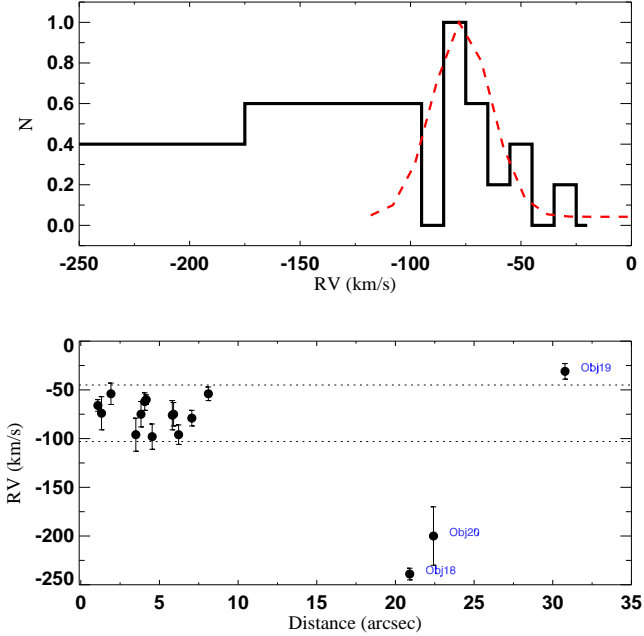}}
\caption{Top: The radial velocity distribution of DBS2003\,179. The line represents the best Gaussian fit. Bottom: 
The radial velocities for our spectroscopic targets as a function of distance from the cluster center. 
The horizontal lines represent the 3$\sigma$ dispersion of the Gaussian fit.  
The error bars represent the random error in determining the RV for each star, the outliers are labeled.}
\label{db179_rvhist}
\end{figure}

The membership of the stars is determined on the basis of the radial velocity 
histogram (see Fig.~\ref{db179_rvhist}). Most of the stars are concentrated 
within the radial velocity interval ($-$100,\,$-$50) 
km\,s$^{-1}$.  The distribution histogram is then fitted with a Gaussian function.
The mean radial velocity as given by Gaussian approximation is $-$74$\pm$16~\,km\,s$^{-1}$.
The uncertainty is determined from Poisson statistics of the measurements and the error of the wavelength solution.
Objects 18 and 19 have very different radial velocities, lie relatively far from the cluster center (Fig.~\ref{db179_rvhist}, bottom panel), and most probably are back- or foreground stars. Object\,20 also has more than 3$\sigma$ of the fit radial velocity and could be field or runway star (see the next paragraph).
 
We defined the following preliminary spectral classification of our targets:

Object\,11 shows broad and complex emission of Br$\gamma$ and N{\sc iii}/O{\sc iii}+He{\sc i} (hereafter 2.116\,$\mu$m complex), while He{\sc ii} 2.188\,$\mu$m is moderately weak in absorption (Fig.~\ref{fg_obj11}) and its spectral classification will be discussed later in this paper. Objects 12, 13, 17, and 19 are quite similar, with the 2.116\,$\mu$m complex in moderately weak emission and Br$\gamma$ and He{\sc ii} in absorption. For Objects 12, 13, and 17 the He{\sc ii} is stronger than  Br$\gamma$. These are typical characteristics of O-type stars not later than O8. Based on their visual $K$-band magnitude, we assigned luminosity class V to Obj\,12, Obj\,17, and Obj\,19 and I/III for Obj\,13. Object\,18 shows only lines in absorption, and the He{\sc ii} line is absent in $K$-band spectra. Our spectral classification is O8-O9, luminosity class V. Object\,14 shows Br$\gamma$ in weak emission, no 2.116\,$\mu$m complex, and He{\sc ii} in $K$-band spectra, but in moderately weak absorption in $H$-band spectra. The object is bright and most probably a supergiant. Objects 4, 15, and 20 (MDM\,32, Mauerhan, Van Dyk \& Morris 2011) show strong emission of Br$\gamma$ and the 2.116\,$\mu$m complex, and the He{\sc ii} is in emission in Obj\,15 and Obj\,20 and in absorption in Obj\,4. Following Crowther et al. (2006) we classified Obj\,4, Obj\,15, and Obj\,20 as Ofpe/WN9, WN9, and WN8-9, respectively. This method of spectral classification, especially in the near-IR is correct within two subtypes, so we adopted this error for our estimates. 

\begin{table*}\tiny
\begin{center}
\caption{EW and radial velocities of the observed stars.}
\label{stars_spectra}
\begin{tabular}{lrrrrrrrrl}
\hline
 & & & & & & & \\[-6pt]
\multicolumn{1}{l}{Object } &
\multicolumn{1}{c}{S/N} &
\multicolumn{1}{c}{HI{11-4}} &
\multicolumn{1}{c}{He{\sc ii}} &
\multicolumn{1}{c}{He{\sc i}} &
\multicolumn{1}{c}{HI{10-4}} &
\multicolumn{1}{c}{N{\sc iii}/O{\sc iii}+He{\sc i}} &
\multicolumn{1}{c}{HI(Br$\gamma$)} &
\multicolumn{1}{c}{He{\sc ii}} &
\multicolumn{1}{c}{RV} \\
\multicolumn{1}{l}{} &
\multicolumn{1}{c}{at 2.134} &
\multicolumn{1}{c}{1.668} &
\multicolumn{1}{c}{1.692} &
\multicolumn{1}{c}{1.702} &
\multicolumn{1}{c}{1.722} &
\multicolumn{1}{c}{2.116} &
\multicolumn{1}{c}{2.166} &
\multicolumn{1}{c}{2.188} &
\multicolumn{1}{c}{km\,s$^{-1}$} \\ 
\hline
& & & & & &  \\[-6pt]
 Obj\,4  & 197    & $-$3.2& 0.8 & $-$7.2& $-$6.4 & $-$6.7& $-$45.2 &1.1   & $-$75$\pm$13\\
 Obj\,11 & 87     & ---   & ---   & ---   & ---    & $-$4.9& $-$10.5 &1.5   & $-$54$\pm$7\\
 Obj\,12 & 78     & ---   & ---   & ---   & ---    & $-$3.7&  1.2  &1.4   & $-$62$\pm$9 \\
 Obj\,13 & 74     & ---   & ---   & ---   & ---    & $-$2.8&  0.7  &1.1   & $-$66$\pm$6 \\
 Obj\,14 & 52     & 0.9 & 1.2 &1.1  & 0.7  & $-$2.3& $-$2.9  &0.6   & \multicolumn{1}{c}{---}         \\
 Obj\,15 & 99     &$-$1.2 &$-$0.9 &$-$1.6 &$-$2.2  & $-$8.0& $-$26.6 &$-$2.2  & $-$96$\pm$17 \\
 Obj\,17 & 68     & 0.9 &0.6  &0.9  & NP   & $-$1.7& 0.5   &0.9   & $-$76$\pm$15\\
 Obj\,18 & 35     & NP  & NP  & 1.9 & 2.1  & NP  & 3.7   &NP    & $-$239$\pm$6 \\
 Obj\,19 & 38     & 0.8 & NP  &1.9  & NP   & $-$2.4& 2.1   &1.1   & $-$31$\pm$8 \\
 Obj\,20 MDM32    & 80  & ---   & ---   & ---    & ---   & $-$20.7 &$-$47.3 &$-$2.9& $-$200$\pm$30\\
\hline
\end{tabular}
\end{center}
\end{table*}

\subsection{Quantitative modeling.}

One of the goals of this work is to better constrain physical parameters for the massive stars. To do so, we proceeded with quantitative modeling of the spectra presented here, as well as of the stars reported in Paper II, which are obtained with the same spectrograph, with similar set-ups and the same reduction. They were separated into two groups. The first containing stars with no evidence of emission at Br$\gamma$, with the second containing the stars that we classified as WR and Of.

As in Hanson et al. (2010), we used {\tt cmfgen} code (Hillier \& Miller 1998). Although the spectra do not show many features, fairly complex atomic models were used in order to treat the line blanketing correctly. In addition we used 306 levels for N{\sc iii} and 343 levels O{\sc iii}, which allowed us to reproduce the N{\sc iii} 2.115 and O{\sc iii} 2.116 $\mu$m blend.  The absence of spectral features in our spectral range does not allow any estimations of the chemical composition of any elements heavier than oxygen so we adopted solar abundances for all them.  It is commonly assumed that the winds are clumpy. We ran models with different clumping factors  but the computed spectra showed that the sensitivity of the spectral features to its change at the resolution of our spectra are below the noise.  The modeling presented in this paper use the same treatment of the clumping as described by Hillier et al. (2003). Following these authors we adopted a value of f=0.1. For the rest of the parameters we adopted the following fitting procedure.

\begin{enumerate}
\item  {\it Velocity law.} The wind velocity was assumed to follow the usual ``$\beta$''-type law.  For the objects with Br$\gamma$ in absorption we used V$_\infty$=2850 km\,s$^{-1}$  and $\beta=0.8$, as in the grid used by Martins et al. (2005) and in our previous paper (Hanson et al. 2010). 

For the emission line stars we used Br$\gamma$  profile to estimate the terminal velocity.  We find that the emission line profiles are not very sensitive to the changes in $\beta$, but the ratio of the intensity of H{\sc i} 1.68 and H{\sc i} 1.74 to Br$\gamma$  are sensitive to it. Higher values of $\beta$ increase H{\sc i} 1.74/Br$\gamma$. Unfortunately, only the spectra of Obj\,4 and Obj\,15 are suitable for such an analysis.  We determine $\beta = 2$ for Obj\,4. The spectrum of Obj\,15 is compatible with $\beta = 1$. We also assumed $\beta$ = 1 for the stars without H spectra or with no emission in  H{\sc i} 1.68 and H{\sc i} 1.74.

\item {\it Temperature and gravity acceleration.} The wind of each model was fitted to a hydrostatic atmosphere with $\log{g}$ and $T_{\rm eff}$, both defined at the radius at which the Rosseland optical depth $\tau_{\rm Ross} = 2/3$. The He{\sc i} 1.70\,$\mu$m, 2.112 and He{\sc ii} 1.69\,$\mu$m and 2.18\,$\mu$m features were used as temperature diagnostics. The gravity acceleration, $\log{g}$, was determined for the stars with absorption lines. In addition to the shape of the wings of the hydrogen and helium lines, we used the intensity of the N{\sc iii} 2.115/O{\sc iii} 2.116 $\mu$m blend. The latter is very sensitive to the changes in $\log{g}$, which makes the determination of the chemical composition of N and O difficult (see next item).  The strength of the He{\sc ii} 2.18$\mu$m line is also sensitive to $\log{g}$. The line became stronger with lower $\log{g}$, which is to be expected. Thus the temperature and $\log{g}$ cannot be determined independently, which increases the uncertainty of the temperature and $\log{g}$. Based on the step of the grid and the correlation between the parameters, we assumed that the temperature is determined within 2000K, although some of the stars allowed for better precision.  The stars that were fitted with models from the grid have an error in $\log{g}$ within 0.25 dex. Objects 15 and 20 have spectra that are not sensitive to $\log{g}$. We adopted values that give reasonable masses for these two stars. The obtained temperatures and gravity accelerations, together with their estimated errors, are given in Table~3, and $\log{g}$ is in cgs-units.

\item {\it Luminosity and reddening.}

We determined the luminosity from the JHK magnitudes.  First we obtained reasonable values for $T_{\rm eff}$ and $\log{g}$ from the fit of the normalized spectra. 
Then we converted the observed magnitudes to fluxes using the Spitzer magnitude-to-flux converter \footnote[1]{http://ssc.spitzer.caltech.edu/warmmission/propkit/pet/magtojy/index.html}.
The equivalent fluxes were calculated from the computed spectra using the 2MASS filters' transmission curves \footnote[2]{http://www.ipac.caltech.edu/2mass}. We then calculated a scaling factor for the adopted luminosity and the reddening for each  object from the least squares comparison of its observed and calculated fluxes. We assumed the Cardelli et al. (1989) reddening law with R=3.1 and fixed the distance to 7.9 kpc.  The derived luminosities are given in Table~3. We estimated the error of 0.2 dex in $\log$(L/L$_\odot)$, taking the errors in temperature, the distance and the reddening into account.

\item {\it Mass loss rate.} Br$\gamma$ was used to determine the mass-loss rate. As for $T_{\rm eff}$ and $\log{g}$, the errors are difficult to determine, due to the dependence of the equivalent width of Br$\gamma$ from several other parameters. We estimated that the error in $\dot{\cal M}/\sqrt{f}$ for the emission line stars should be less than 20\%. The stars with  Br$\gamma$ in absorption were modeled with the adopted mass-loss rate. Once the luminosity of the object was determined, the values of $\dot{\cal M}$ used in its model were scaled using Eq. (4) from Morisset \& Georgiev
(2009) to the transformed radius (Schmutz et al., 1989), and therefore the strength of the computed spectral lines is conserved for the determined luminosity. The results are given in Table~3.

\item {\it Chemical composition.} The ratio between the He and H lines in both the $H$ and $K$ spectra were used to fix He/H ratio. The N{\sc iii} 2.115/O{\sc iii} 2.116 $\mu$m blend was used to estimate the nitrogen and oxygen abundances.  As mentioned above, the blend is more sensitive to changes in $\log{g}$ than to the changes in the chemical composition, so the abundances determined based only on $K$ spectrum and only on that spectral feature should be taken as rough estimates. Our analysis shows that, for all absorption line stars, a solar oxygen and nitrogen abundances is acceptable.  The C{\sc iv} 2.07 $\mu$m line is present at the blue edge of our low-resolution spectra (Hanson et al. 2010). Although this line could be used to estimate the carbon abundance, in most of the cases, it is in the beginning of the ISAAC spectral range, is thus very noisy, and does not allow any rigorous analysis. In addition, the line has the same high sensitivity to $\log{g}$ as the N{\sc iii} 2.115/O{\sc iii} 2.116 $\mu$m blend, and it is also sensitive to the temperature.  In most of the absorption line stars, the observed C{\sc iv} 2.07 is compatible with solar carbon abundance models except for Obj\,5, for which the model predicts an observable line, but none is detected.

The stronger winds of the emission line stars make their spectra less sensitive to the underlying photosphere and therefore to $\log{g}$. The N{\sc iii} 2.115 line is strong and dominates the  N{\sc iii} 2.115/O{\sc iii} 2.116 $\mu$m blend. We still kept the oxygen abundance for these stars to its solar values, but we determined the nitrogen abundance, which in all cases is increased by a factor of 5-10 with respect to the solar. The C{\sc iv} 2.07 is detectable in the spectra of the Obj\,3, Obj\,4, and Obj\,8. Our current model for Obj\,3 is for its ``active'' phase (see below) based on our new spectrum, which does not cover the C{\sc iv} 2.07 region. The C{\sc iv} 2.07 is detected in the spectrum of its ``passive'' phase. If this star is Of?p (see the discussion), the carbon lines are expected to vary, so the current model cannot be applied to its old spectrum, and we cannot derive the carbon abundance. Both Obj\,4 and Obj\,8 show much weaker C{\sc iv} 2.07 than the model with solar carbon abundance. To reconcile the observed and model spectra, we reduced the carbon abundance of the models.  The depleted carbon abundance and the increased nitrogen abundance are more evidence of the advanced evolution of these stars.
\end{enumerate}

The results are shown in Table~\ref{tab_model_par} and model spectra are compared with those observed in Figs.~\ref{db179_abs} and \ref{db179_emi}. After analyzing all the information presented above we made the following conclusions for the most massive stars in DBS2003\,179.

\begin{table*}\small
\begin{center}
\caption{Model parameters of the stars with spectra.} 
\label{tab_model_par} 
\begin{tabular}{lllrrrrrrr} 
\hline
 & & & & & & & & & \\[-6pt]
\multicolumn{1}{l}{Object} &
\multicolumn{1}{l} {Spec. type} &
\multicolumn{1}{l}{$T_{\rm eff}$} &
\multicolumn{1}{l}{$\log\,{g}$} &
\multicolumn{1}{l}{${log(L/L_\odot})$ } &
\multicolumn{1}{l}{$\dot{\cal M}/\sqrt{f}$} &
\multicolumn{1}{l}{V$_\infty$} &
\multicolumn{1}{l}{$\beta$} &
\multicolumn{1}{l}{He/H} &
\multicolumn{1}{l}{log [N/H]}\\
\multicolumn{1}{l}{} &
\multicolumn{1}{l}{} &
\multicolumn{1}{l}{[kK]} &
\multicolumn{1}{l}{cgs-units} &
\multicolumn{1}{l}{} &
\multicolumn{1}{l}{$M_\odot$/yr } &
\multicolumn{1}{l}{$km\,s^{-1}$} &
\multicolumn{1}{l}{} &
\multicolumn{1}{l}{} &
\multicolumn{1}{l}{}\\ 
\hline
 & & & & & & & & & \\[-6pt]  
Obj\,1$^c$ &O5/6V	       &35.0 $\pm$2.5   &  3.50$\pm$0.25 & 5.54$\pm$0.2     & 1.81E-05$^b$ &   2825$^a$       &   0.8 &   0.1 &   -0.1 \\
Obj\,2$^c$ &O5/6V         &32.5 $\pm$2.5   &  3.25$\pm$0.25 & 5.24$\pm$0.2    & 1.07E-05$^b$ &   2825$^a$       &   0.8 &   0.1 &    0.0 \\
Obj\,3$^d$ &O5/6If        &30.0 $\pm$1.25  &  3.10$\pm$0.12 & 5.52$\pm$0.2    & 8.58E-06     &    800$\pm$50    &   1.0 & 0.25$\pm$0.025 &    0.7$\pm$0.1 \\
Obj\,4$^d$ &Ofpe/WN9      &30.0 $\pm$1.25  &  2.90$\pm$0.12 & 6.15$\pm$0.2    & 3.34E-05     &    500$\pm$50    &   1.9 &   0.1 &    1.0$\pm$0.1 \\
Obj\,5$^c$ &O5/6I         &35.0 $\pm$2.5   &  3.25$\pm$0.25 & 5.42$\pm$0.2    & 1.47E-05$^b$ &   2825$^a$       &   0.8 &   0.1 &    0.0 \\
Obj\,6$^d$ &YSO?          &32.5 $\pm$5.0   &  3.50$\pm$0.50 & 4.62$\pm$0.5    & 3.22E-07     &    300 $\pm$50   &   1.0 &   0.1 &    0.0 \\
Obj\,8$^d$ &O4/5I         &34.0 $\pm$1.5   &  3.10$\pm$0.15 & 5.62$\pm$0.2    & 8.27E-06     &   1250$\pm$50    &   1.0 & 1.0$\pm$0.025  &    0.6$\pm$0.2 \\
Obj\,10$^c$ &O8V           &33.5 $\pm$1.5   &  3.00$\pm$0.3 & 4.98$\pm$0.2    & 6.26E-06$^b$ &   2270$^a$       &   0.8 &   0.1 &   -0.1 \\
Obj\,12$^c$ &O4/5V         &35.0 $\pm$2.5   &  3.50$\pm$0.25 & 5.47$\pm$0.2   & 1.60E-05$^b$ &   2825$^a$       &   0.8 &   0.1    &   -0.1 \\
Obj\,13$^d$ &O4III         &34.0 $\pm$1.25  &  3.10$\pm$0.12 & 5.76$\pm$0.2   & 1.05E-05     &   1250$\pm$50    &   1.0 &   1.0    &    0.6$\pm$0.1 \\
Obj\,14$^d$ &O5/6If        &32.5 $\pm$2.0   &  3.50$\pm$0.30 & 5.72$\pm$0.2   & 1.18E-06     &    300$\pm$50    &   1.0 &   0.1      &    0.0 \\
Obj\,15$^d$ &WN8-9h        &37.0 $\pm$1.0   &  3.60$\pm$0.10 & 6.29$\pm$0.2   & 4.05E-05     &   1000$\pm$50    &   1.0 & 0.15$\pm$0.025  &    1.1$\pm$0.1 \\
Obj\,17$^c$ &O6/7V         &35.0 $\pm$2.5   &  3.50$\pm$0.25 & 5.35$\pm$0.2   & 1.29E-05$^b$ &   2825$^a$       &   0.8 &   0.1      &   -0.1 \\
Obj\,18$^c$ &O8/9V         &35.0 $\pm$2.5   &  3.50$\pm$0.25 & 4.96$\pm$0.2   & 2.18E-06$^b$ &   2400$^a$       &   0.8 &   0.1    &   -0.1 \\
Obj\,19$^c$ &O5/6V         &35.0 $\pm$2.5   &  3.50$\pm$0.25 & 5.26$\pm$0.2   & 1.11E-05$^b$ &   2825$^a$       &   0.8 &   0.1    &   -0.1 \\
Obj\,20$^d$ &WN8-9         &35.0 $\pm$2.5   &  3.40$\pm$0.25 & 6.23$\pm$0.2   & 1.18E-04     &   1250$\pm$50    &   1.0 & 1.0$\pm$0.025  &    1.0$\pm$0.2 \\
\hline
\end{tabular}
\end{center}
{\footnotesize $^a$ Value adopted in the grid of models.} {\footnotesize $^b$ Value scaled from the adopted in the grid to conserve the transformed radius (see text).} {\footnotesize $^c$ Models selected from the grid.} {\footnotesize $^d$ Models calculated to fit the particular spectrum.} \end{table*}

\begin{figure}

\resizebox{\hsize}{!}{\includegraphics{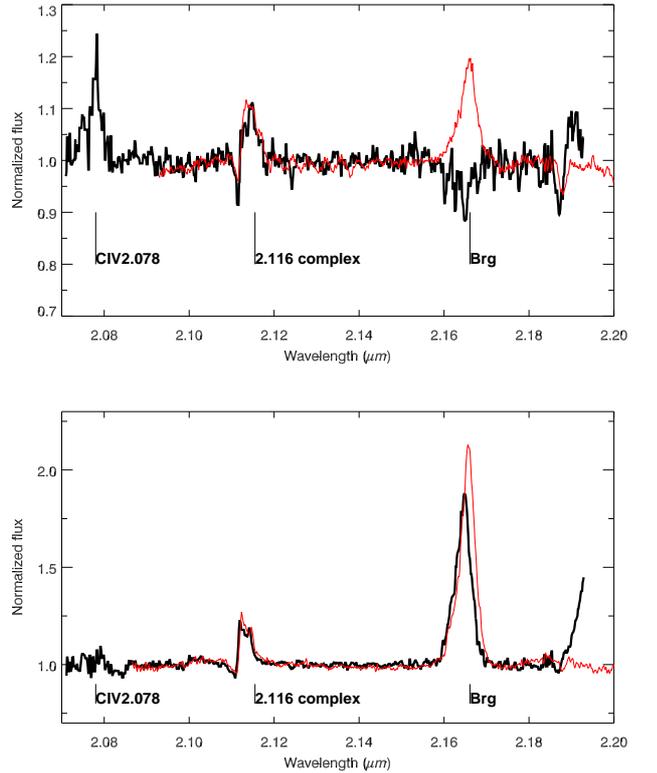}}
\caption{Upper panel: ISAAC, VLT, ESO, $K$-band spectra of Obj\,3 taken in 2007 (thick black line) and 2008 (thin red line).
Lower panel: The same but for Obj\,4.}
\label{comp_obj3_4}
\end{figure}

Object\,3 shows strong emission in Br$\gamma$ and the 2.115/2.116\,$\mu$m complex, while the He{\sc ii} is in absorption.
With $T_{\rm eff}$=30000\,K, it does not look significantly different from many mid-O extreme stars, with an extended atmosphere, and it has most probably early O5/6If spectral type. Since the star was observed twice (in our 2007 and 2008 ISAAC runs), we compared its $K$-band spectra and they showed significant differences. For example, in our 2008 ISAAC run the Br$\gamma$ line is in emission (Fig.~\ref{comp_obj3_4}, upper panel); as opposed to the 2007 run, where it is in absorption), some variations in the 2.116\,$\mu$m complex can also be seen. If confirmed, it would be an interesting case of a star in transition. The star resembles HD\,191612 (Walborn \& Howarth 2007) and the class of Of?p stars (see Naz\'e et al. 2010 and references therein), which show similar variations in the optical bands. The Of?p phenomenon covers stars presenting spectral variations (in the Balmer, HeI, CIII, SiIII lines), strong CIII  ($\lambda$ 4650\,\AA) in emission (at least in some phases), narrow P Cygni or emission components in the Balmer hydrogen lines and He I lines (in some phases), and ultraviolet wind lines weaker than for the typical Of supergiants. Object\,3 has increased He/H=0.25 and N/H=5 times solar (by mass), which closely resembles the composition of HD108 (Martins et al. 2010). The two stars have similar luminosities ($\log{L/L_\odot}$ = 5.7 for HD108 and $\log{L/L_\odot}$ = 5.5 for Obj\,3), and although Obj\,3 is cooler, the temperatures are compatible within the errors.  It also has a peculiar position in the reddening-free color-magnitude diagram (see next paragraph). Thus we conclude that star Obj\,3 is a candidate for another member of the Of?p class (see Walborn et al. 2010, for the recent review of the five known members in the Milky Way of this type) however, long-term monitoring is needed to confirm this hypothesis.

\begin{figure}
\resizebox{\hsize}{!}{\includegraphics{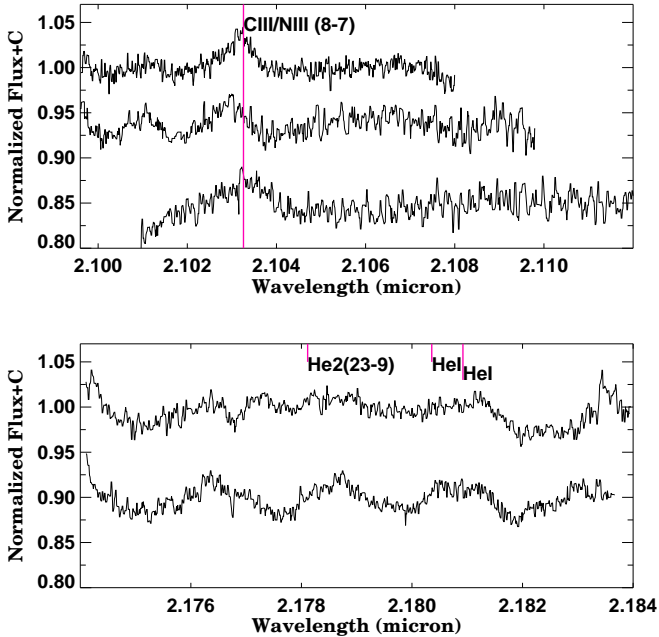}}
\caption{Phoenix spectra of Object\,4 of DBS2003\,179.} \label{db179_gemini} \end{figure}

Object\,4, classified as Ofpe/WN9 has a model temperature of $T_{\rm eff}$=30000\,K, but the He/H ratio is low (0.1), and it shows a very high luminosity. It was also observed twice. As can be seen in Fig.~\ref{comp_obj3_4}, the 2.116\,$\mu$m complex does not change significantly (EW$_{2007}$=$-$6.0 vs. EW$_{2008}$=$-$6.7), but in 2008 the Br$\gamma$ line is stronger with EW$_{2008}$=$-$45.2 vs. EW$_{2007}$=$-$37.2. This variability, if confirmed, could indicate presence of a close binary companion. Object\,4, however is placed very close to Obj\,15, which was also classified as a WN8 star, and with 0.6 wide slit Obj\,15 could have contaminated the spectrum. To check this, we searched for additional diagnostic lines and obtained high-resolution (R$\approx$50000) Phoenix, Gemini spectra in the two  spectral settings centered on 2.104 and 2.179 $\mu$m, respectively. In general, these spectral settings  contain the following lines for the hot stars:  HeI 2.1120\,$\mu$m, CIII/NIII (8-7) 2.1032\,$\mu$m, HeII(23-9) 2.1786\,$\mu$m,  HeI 2.1809\,$\mu$m, and  HeI 2.1814\,$\mu$m. In the upper panel of Fig.~\ref{db179_gemini} are plotted three spectra that were taken on 27 May, 13 June, and 24 June 2009, and centered on 2.104\,$\mu$m.  The lower panel shows the 2.179\,$\mu$m setting taken on 4 June and 14 June 2009. The spectra are corrected with the mean radial velocity of the cluster. As can be seen from the figure, the CIII/NIII (8-7) lines at 2.1038\,$\mu$m are well defined and clearly present in emission. Moreover, the center and the shape of the line seem to be variable. The Julian date of the observation, center of the line (as measured from Gaussian approximation of the line profile by {\tt splot} routine in IRAF), and the relative radial velocity with respect to the first spectrum and its error (measured by {\tt fxcor} in IRAF) are given in Table~\ref{obj4_gemini}.  The width of the slit is small (0.34 arcsec) and is specially centered to avoid the influence of the nearby Object\,15. Thus, most probably Object\,4 is a binary system. Assuming this, the relatively small variability of  Br$\gamma$ and the good fit of the spectrum with a single star model point to a relatively small contribution by the secondary to the observed spectrum. Therefore one can conclude that the parameters derived from the spectra are representative of the brighter star in the system, but more data are needed to make more definitive conclusions.

\begin{table}
\begin{center}
\caption{Relative radial velocities of CIII/NIII line in Object\,4.} 
\label{obj4_gemini} 
\begin{tabular}{cccc} 
\hline 
\multicolumn{1}{l}{JD} & \multicolumn{1}{c}{Central Wavelength} & \multicolumn{1}{c}{RV} & \multicolumn{1}{c}{$\sigma_{RV}$} \\ \multicolumn{1}{l}{Day} & \multicolumn{1}{c}{$\mu$m} & \multicolumn{1}{c}{km\,s$^{-1}$} & \multicolumn{1}{c}{} \\ 
\hline
54978.61149  & 2.102911 & 0   & 4 \\
54995.77851  & 2.103142 & -42 & 4\\
55006.82398  & 2.103249 & -62 & 5 \\
\hline
\end{tabular}
\end{center}
\end{table}

Object\,5 was classified in Paper II as O-type dwarf star.  Considering, however that the physical parameters derived for this star from the models are more compatible with the supergiants (even though under luminous) and its peculiar position on the color-magnitude diagram, we reclassify the star as O5/6I.  

Object\,6 has a $K$ spectrum with only Br$\gamma$ in emission, but the line shows two components. One is broad, corresponding to a stellar wind, and another much narrower which might originate 
in a circumstellar shell or another ionized gas on the line of sight. We tentatively classified this object as a YSO, but based only on one line, it is only a rough estimate. In the same way, we derived a temperature for the star, such that the computed spectrum has He{\sc i} 2.112 and He{\sc ii} 2.18 lines weaker than the observed noise, and the mass loss rate and V$_\infty$ fits Br$\gamma$ profile, but the values presented in Table~3 are only estimates. We reflected that by assigning larger error bars to its parameters.  

Objects\,8  and 13 have a very uncommon spectrum. The spectra show a strong He{\sc i}/N{\sc iii}/O{\sc iii} complex with weak He{\sc ii} 2.18 absorption line and very weak Br$\gamma$. We reproduced this spectrum with a helium and nitrogen-rich wind with depleted carbon.  This makes the stars evolved, but their mass-loss rates are very low, which makes the objects unusual. Unfortunately, we only have $K$-band spectra for these objects, and the obtained set of parameters might not be unique. Other solutions with different chemical composition might be possible. In any case, the stars have increased nitrogen abundance and are evolved from the main sequence.

\begin{figure}
\resizebox{\hsize}{!}{\includegraphics{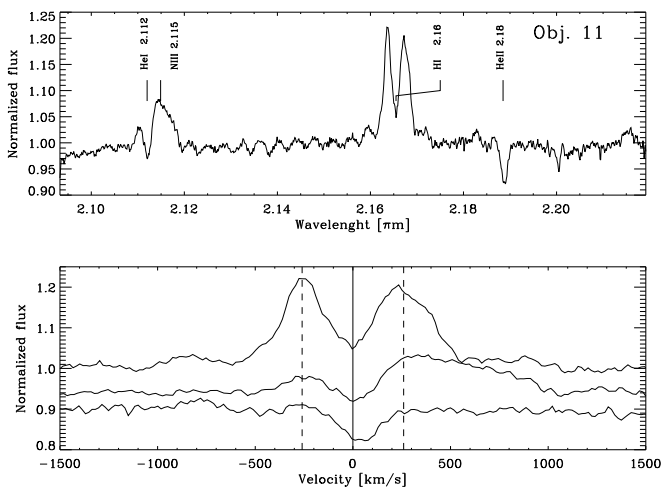}}
\caption{Upper panel: K spectrum of Object\,11.
Lower panel: from top to bottom Br$\gamma$, HeI 2.11 and HeII 2.18 in velocity space. The lines are shifted by $-$70 km\,s$^{-1}$ so the center of Br$\gamma$ absorption component is at 0 km/s. The two vertical dashed lines are at $\pm$ 270 km/s and mark the position of Br$\gamma$ emission components. Notice that He{\sc ii} 2.18 is narrower than Br$\gamma$.} 
\label{fg_obj11} 
\end{figure}

Object\,11 has a very peculiar spectrum (Fig.~\ref{fg_obj11}, upper panel). The spectrum shows a double peak Br$\gamma$, a P Cyg like He{\sc i} 2.11, and absorption in He{\sc ii} 2.18. It is tempting to interpret the symmetric Br$\gamma$ profile as the result of rapid rotation because the split between the two peaks is 540 km\,s$^{-1}$, equivalent to 270 km\,s$^{-1}$ rotation. But a close inspection of the other two lines does not support this hypothesis. The lower panel of Fig.~\ref{fg_obj11} shows the profiles of the three lines in the velocity space. It is obvious that He{\sc i} 2.11 profile is not P Cyg, but rather a composition of an emission line and an absorption component, centered on the same velocity as the Br$\gamma$ central feature. On the other hand, He{\sc ii} 2.18 is redshifted by $\sim$40 km\,s$^{-1}$, and its profile is narrower than what one might expect for a 270 km\,s$^{-1}$ rotational velocity (Fig.~\ref{fg_obj11}, lower panel). Based on that, one can suggest that the star is a binary with one component similar to Obj\,4 (Br$\gamma$ and He{\sc i} 2.11 in emission and He{\sc ii} 2.18 in absorption) and the other a cooler secondary with deep Br$\gamma$ and He{\sc i}. We estimated $L$=4.34\,10$^5$\,$L_\odot$ using Obj\,4's model continuum as a reference and A$_V$=21.5. This makes the star not particularly bright, which is an argument against the binary hypothesis. It seems that the star is peculiar and its interpretation requires spectra in wider wavelength range.

Object\,15 is classified as WN9, has a model temperature of $T_{\rm eff}$ =37000\,K, but the He/H ratio is low (0.15), and it shows a very high luminosity. This could indicate that the star is not a ``classical'' WR star, but rather a star in the transition O$\rightarrow$WN$\rightarrow$LBV. Martins et al. (2008) found similar composition for the stars B1, F1, F6, and F9 of the Arches cluster. They classified these stars as WN8-9h even the composition resembles the one of a O star rather than the WR star. The authors conclude that the stars are descendants of the stars that are more massive than 60 M$_\odot$. The same conclusion can be applied to Obj\,4 and Obj\,15 of DBS2003\,179.  

Our model of Obj\,20 is the most unsatisfactory. We were not able to reproduce the peculiar shape of Br$\gamma$. The closest model has an increased He/H and N/H, high-mass loss rate, and  temperature adequate for a WN star. But the profile of Br$\gamma$  has a peculiar rectangular shape which might be a result of a binary with two emission line components.
The high luminosity, the temperature, and the chemical composition of this star are similar to the parameters of HD5980 (Georgiev et al. 2011). One can speculate that Obj\,20 might manifest similar unstable behavior, and it needs further monitoring.

\subsection{Relative proper motions.}

\begin{figure}
\resizebox{\hsize}{!}{\includegraphics{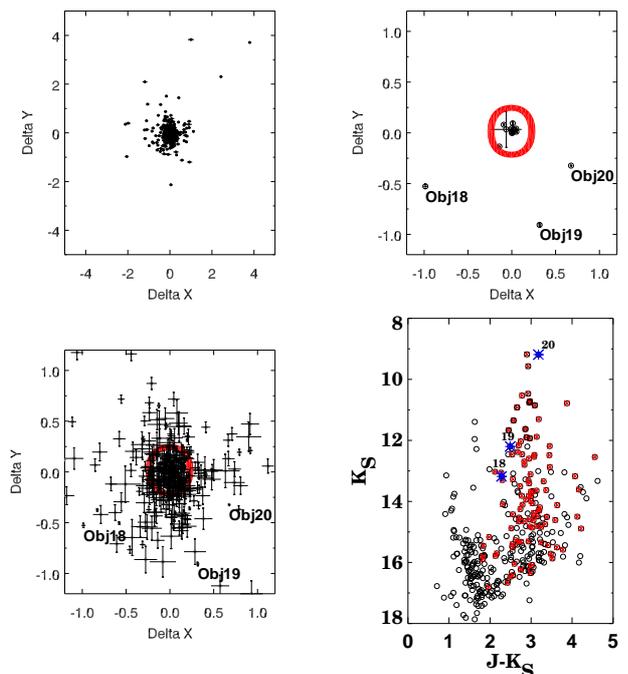}}
\caption{Top panels: Vector point diagrams of displacement of all stars within 35 arcsec of the cluster center (left) and spectroscopically confirmed cluster members (right) in units of ISAAC pixels (185 mas/pixel) after 5 years. Bottom panels: A zoom of the central 2$\times$2 pixel vector point area. The most probable cluster members are within the red circle with 0.25 pixels radius (left). They are marked with red crosses on the color-magnitude diagram of this zoomed area (right). All plots show only stars with root-mean-square positions $<$0.08 pixels in each coordinate, the Objs\,18, 19 and 20 are also labeled.
}
\label{db179_pm}
\end{figure}

Objects\,18, 19, and 20 are situated at the limit of the 32\,arcsec cluster radius determined in Paper II (see Fig.~\ref{db179_chart}) with radial distances from the center of 20, 30, and 17 arcsec, respectively, and thus they need special attention in order to verify their membership. We calculated the relative proper motions of all stars within 35 arcsec of the cluster center. This small area was chosen in order to minimize the effects of distortion errors. We used the Anderson et al. (2006) method, which calculates the relative displacement between two epochs frames with respect to a local reference system of common stars. The best intrinsic astrometric accuracy of 0.006-0.007 pixels in x and y coordinates measured by PSF photometry (the {\tt CENTER} task) was obtained for a PANIC Br\,$\gamma$ image taken in 2003 and an ISAAC $K_{\rm S}$-band image taken in 2008, which we chose as first- and second- epoch frames. At the distance of DBS2003\,179 a five-year baseline is not enough to measure the proper motion of the cluster stars from the ground. Instead, our intent was to track the possible high proper motion stars relative to our reference stars. The PPMXL catalog of positions and proper motions (Roeeser  et al. 2010) lists 40 objects within a 1\,arcmin radius centered on the cluster. We chose those 25 of them with the smallest proper motions ($<$0.07 pixels in 5 years) to create our local reference system. The linear transformation between the two frames was computed using {\tt GEOMAP} in IRAF, considering the shift, rotation, and scaling factor.  Finally, we computed a global transformation-based displacement for each star defined as the difference between the Frame 1 position and the Frame 1 position implied from Frame\,2 (based on the positions of the common stars). The calculated errors were associated with centroid measurements for each star and the root-mean-square errors associated with the transformation. We then plotted the vector-point diagram (Fig.~\ref{db179_pm}, top-left panel) of the displacement of all stars within 35 arcsec of the cluster center, and the spectroscopically confirmed cluster members (top-right panel). On the basis of spectroscopically confirmed members, and  after taking their errors into account, we outlined a circle with a 0.25\,pixel radius that isolated the cluster and field stars. The stars within this circle were considered the most probable cluster members (bottom-left panel). The stars Obj\,18, Obj\,19, and Obj\,20 are notably outside this zone. In the color-magnitude diagram of the two zones shown in the bottom right-hand panel of Fig.~\ref{db179_pm} (open circles are field stars, red crosses represent cluster members while Objs\, 18, 19, and 20 are blue asterisks) Obj\,20 falls exactly in the WR zone of the cluster CMD (together with Objs\,4 and 15), while Obj\,19 is at the limit and Obj\,18 is more than 3 $\sigma$ from the mean cluster line. 

\begin{figure}
\resizebox{\hsize}{!}{\includegraphics{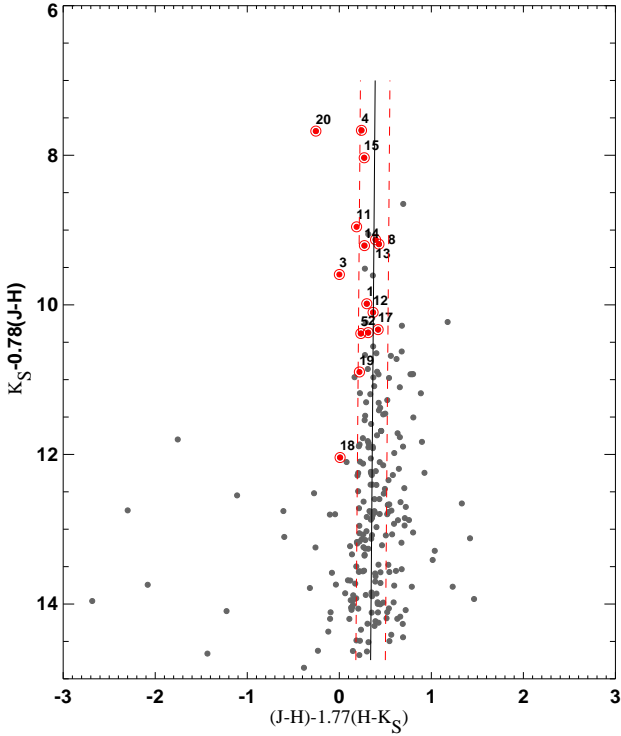}}\hspace{1cm}
\resizebox{\hsize}{!}{\includegraphics{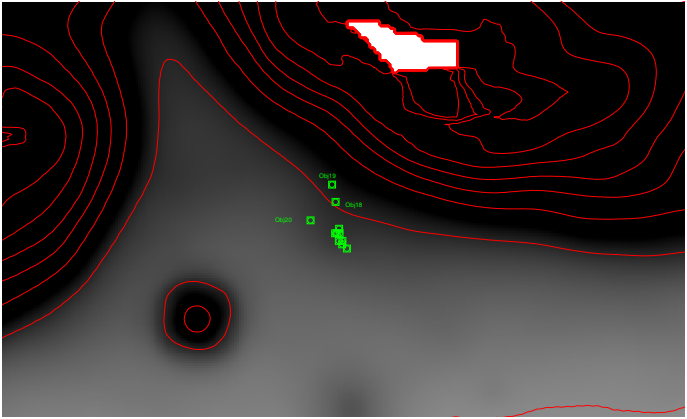}}
\caption[]{Upper panel: The reddening free color of $(J-$$H)$$-1.77(H-$$K_{\rm S})$ vs. $K_{\rm S}$$-0.78(J-$$H)$ magnitude. The solid line is a mean cluster line, while the dashed lines represent 3$\sigma$ deviation. Bottom panel: WISE 22$\mu$m image with density contours overplotted. The stars with spectra are labeled. 
}
\label{db179cmd_isaac_red_free}
\end{figure}

The spectroscopically calculated reddening of $A_{V}$=14.5 for Obj\,18 and 15.4 mag for Obj\,19 are lower than the median value ($A_{V}$=18) of the rest of the spectroscopic targets, while Obj\,20 shows no differences. The last check we performed for these three stars was to note their position on the reddening-free color-magnitude diagram in order to estimate the effect of differential reddening. We used the reddening-free parameter $Q=$$(J-$$H)$$-$1.77$(H-$$K_{S})$, as defined by Negueruela et al. (2007) for OB stars. Figure~\ref{db179cmd_isaac_red_free} shows this reddening-free parameter vs. a corrected $K_{\rm S}$ magnitude. Object\,20 has a peculiar position, which could be due to local dust overdensity and/or dust production. The first possibility is unlikely if taking the position of the star on the WISE 22$\mu$m image into account (Fig.~\ref{db179cmd_isaac_red_free}). Object\,18 is far from the mean cluster line, while Obj\,3 also has a rather peculiar position.

After combining these results with the radial velocity measurements, we conclude that Obj\,20 most probably is a cluster member, although its runaway status cannot be excluded, while Objs\,18 and 19 are field stars.

\section{Cluster parameters} 

Mauerhan, Van Dyk \& Morris (2011) used the calibration photometry values for WN8-9 stars from Crowther et al. (2006), and assumed Obj\,20 (MDM\,32) as a cluster member to calculate a distance to DBS2003\,179 of 6.4\,kpc with the associated error of 0.5-1\,mag. In contrast, Davies et al. (2012) assumed that the cluster is physically associated to a nearby molecular cloud and its massive YSOs to derive a far-side kinematic distance of 9 kpc.  Using the same technique described in Paper II and the intrinsic color and absolute magnitudes from Martins \& Plez (2006), we recalculated the distance to the cluster. The average value of stars Objs.\,1, 2, 3, 8, 10, 12, 13, 14, and 17 gives distance to the cluster ($m$$-$$M$)$_0$=14.5$\pm0.3$\,mag or 7.9$\pm0.8$\,kpc, in excellent agreement with the result obtained in Paper II.  We did not consider the WR stars (Objs.\,4, 15, and 20) or the stars with very peculiar behavior (Objs.\,5, 6, and 11). The error is calculated from a standard deviation of the mean value, quadratically added errors from photometry and the variance due to an uncertain spectral classification. It is comparable within the errors with the kinematical distance reported by Davies et al. (2012), especially considering that, in general, the kinematically determined distances are longer than those obtained from spectroscopic parallaxes. Thus the cluster is very close to the galactic center and could be on the other side of the galactic center.

\begin{figure}
\resizebox{\hsize}{9cm}{\includegraphics{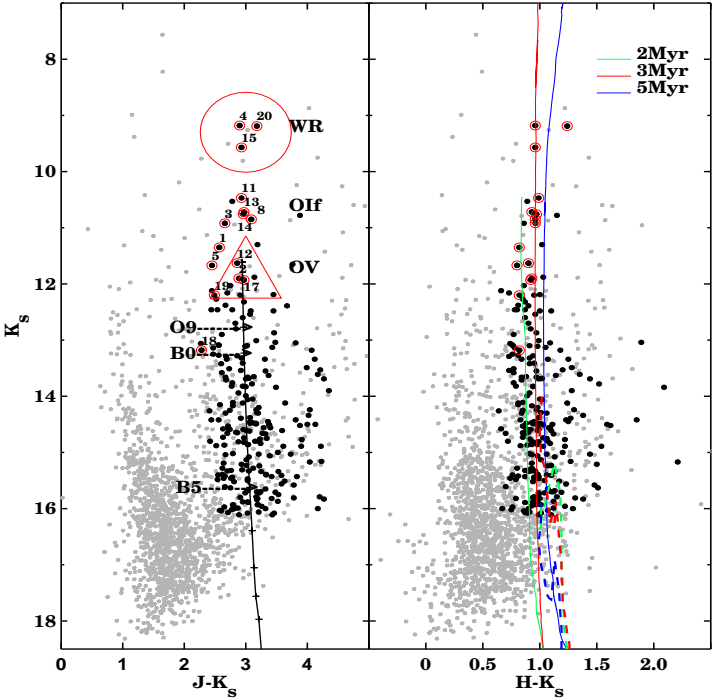}}
\caption{Left panel: The $K$$_{\rm S}$ vs. ($J-$$K_{\rm S}$) color-magnitude diagram of DBS2003\,179. All stars in the field of view are shown as light circles, while the cluster members are plotted with bold circles. The theoretical main sequence from Schmidt-Kaler (1982) is drawn with a solid line corrected for the adopted distance modulus of ($m$$-$$M$)$_0$=14.5\,mag and $E(B$$-$$V)$=5.4. Stars with known spectral types are labeled. The large circle and the triangle schematically mark the zones of WR and OV class stars, respectively.
Right panel: $K_{\rm S}$ vs. ($H-$$K_{\rm S}$) color-magnitude diagram. The solar-metallicity Geneva isochrones (solid lines) of 2(green), 3(red), and 5(blue) Myrs (Lejeune \& Schaerer 2001) and the corresponding PMS isochrones (dashed lines) from Siess, Dufour, \& Forestini (2000) are shown.
}
\label{db179cmd_isaac}
\end{figure}

The $K$$_{\rm S}$ vs. $(J-$$K_{\rm S})$ and $K_{\rm S}$ vs. $(H-$$K_{\rm S})$ color-magnitude diagrams of DBS2003\,179 are plotted in Fig.~\ref{db179cmd_isaac}.  The potential cluster members (N=230) are selected by statistical decontamination as described in Paper I and the proper motion analysis described above. The theoretical main sequence (Schmidt-Kaler 1982) is shown and is corrected for the adopted $E(B$$-$$V)$=5.4 and
($m$$-$$M$)$_0$=14.5$\pm$0.3\,mag. The stars with known spectral type are labeled, the zones of WR and OV class stars, respectively. In general, the $K_{\rm S}$ magnitudes of the objects are consistent with the luminosity class obtained from  the spectral classification.  Object\,5 is slightly underluminous for its classification as a supergiant.

The age can be estimated by fitting the observed CMD with solar-metallicity Geneva isochrones (Lejeune \& Schaerer 2001) in combination with pre-main sequence (PMS) isochrones (Siess, Dufour \& Forestini 2000).  Starting with the isochrones set to the previously determined distance modulus and reddening, we apply shifts in magnitude and color until the fitting statistics reach a minimum value (i.e. difference in magnitude and color of the stars from the isochrone should be minimal) for both the main sequence and PMS isochrones. We get an ``age envelope'' of 2-5~Myr.  Figure\,\ref{db179cmd_isaac} shows the isochrone fits superimposed on the decontaminated CMD.  As can be seen in the figure, however, for such a young star cluster, the main sequence isochrones are almost vertical lines in near-infrared, and it is hard to determine the precise age even when using the PMS isochrone set, especially in this case, when the photometry is not deep enough. For this reason we determine the age using the HR diagram for the most luminous stars in the cluster by comparing the distribution of the stars with the latest isochrones and stellar tracks (with rotation) taken from  Ekstr{\"o}m et al. (2012). The temperature and luminosities are taken directly from our quantitative modeling (see Table~\ref{tab_model_par}). The WR stars seem to fit the 3 Myr isochrone well, while the rest of the stars seem to group around 5 Myr. A similar spread in age between O stars and WR stars was reported by Martins et al. (2008) for the Arches cluster and Liermann et al. (2012) for the Quintuplet cluster. They found ages of 2 to 3~Myr for the most luminous WN stars of Arches, while the O stars cover a range of 2 to 4~Myr. They conclude that this might be because the most massive stars formed last in the cluster. Similar arguments are applied by Liermann et al. (2012) for the  difference in age of the OB and WN stars in the Quintuplet. Thus, we adopt the age of the massive star population of 4$\pm$0.7 Myr for DBS2003\,179. The stars, which show possible binary components, are not taken into consideration.  The comparison with the isochrones from Girardi et al. (2002, ``Padova isochrones'') corresponds well with the derived from Geneva set "age envelope". The error of this method can be estimated as $\pm2$ Myr taking into account the photometric errors, errors of model fits, distance and reddening determinations.   


\begin{figure}[h]
\resizebox{9cm}{!}{\includegraphics{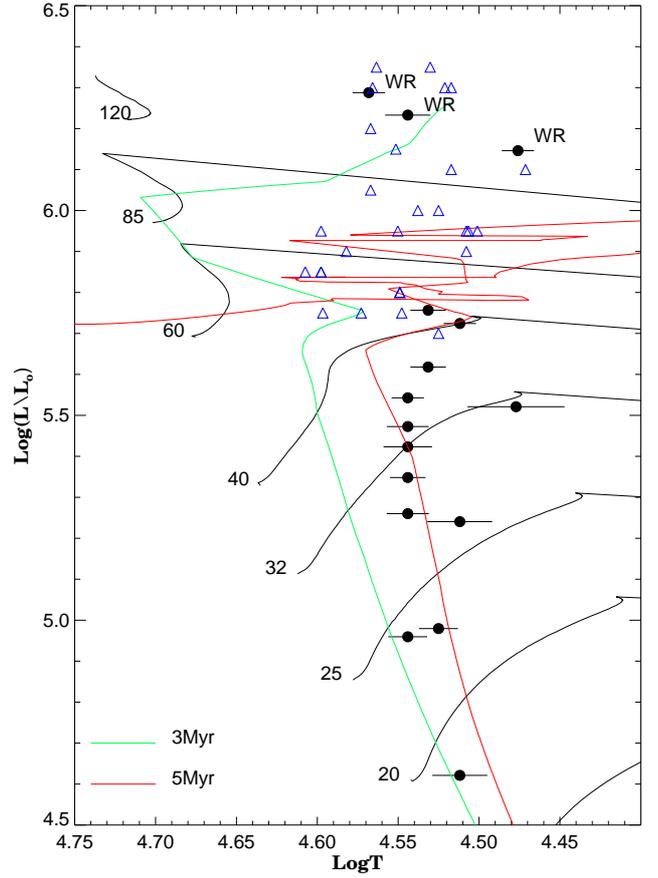}}
\caption{Hertzsprung-Russell diagram for DBS2003\,179 (circles) and Arches (open diamonds, Martins et al. 2008) with evolutionary tracks from Ekstr{\"o}m et al. (2012) overplotted. The solid horizontal lines represent the mass tracks for stellar masses 120, 85, 60, 40, 32, 25, and 20\,$\cal M_{\odot}$.  The vertical solid lines represent 3 and 5 Myr stellar isochrones from the Geneva library. } 
\label{hr} 
\end{figure}

Luminosity functions (LFs) were created for the cluster in bins of 0.5 $K$-band magnitudes.  To calculate the completeness for each magnitude bin, we created artificial stars within the given magnitude ranges on the cluster image.  The completeness was calculated by finding the recovery fraction of artificial stars per magnitude, using the same detection methodology we employed for our PSF photometry (see Sect. \ref{obs_data_red}).

\begin{figure}[!h]
\resizebox{9cm}{!}{\includegraphics[angle=90]{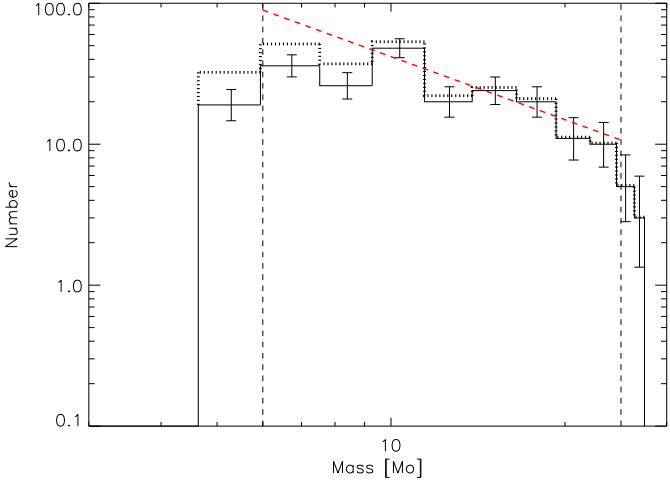}}
\caption{IMF of DBS2003\,179.  The slope is calculated between the mass limits denoted by vertical dashed lines.  The completeness-corrected values are shown as a dotted histogram.  Overplotted as a red dashed line is the least-squares fitted slope (based on the completeness-corrected histogram).}
\label{imf}
\end{figure}

The LF was converted to the initial mass function (IMF) using the 4 Myr isochrone of Geneva assuming a solar metallicity.  Since our photometry does not extend deep enough to detect pre main sequence stars, they were not a concern, and the stars within 5\,$\sigma$ around isochrone were considered to be main sequence. The WR stars were also not considered in calculating the IMF's slope. 

Figure \ref{imf} shows the IMF for DBS2003\,179.  The slope was calculated over the mass range limited by the vertical dashed lines.  The upper mass limit was set to where the IMF slope starts to skew due to the effects of saturated stars, with the lower limit set to where the effects of incompleteness affects the slope.  Completeness-corrected IMF is shown with dotted line.  Uncertainties are also shown and were calculated from the small-number stochastic uncertainties from Gehrels (1986), where the upper limit is given by $1+\sqrt{d\rm{N}+0.75}$ and the lower limit is given by $\sqrt{d\rm{N}-0.25}$. In our case, $d\rm{N}$ is simply the number of objects within a given interval.  The Gehrels equations give an approximation of higher and lower limits for each bin for low number statistics.  Applying the equations to each bin gave the uncertainties.

Our calculated slope of $\Gamma$=$-$1.49$\pm$0.35, with uncertainties that propagate from those on the individual mass bins, is consistent with the Salpeter slope ($\Gamma$=$-$1.35).  This slope was used to extrapolate the masses of the observed main sequence stars to provide a total mass for the cluster.  For extrapolation of masses below $<$0.5\,$\cal M$$\odot$ we used the Kroupa (1998) value of $\Gamma$=$-$0.3.  This gives a total mass of 25300$^{+8100}_{-7500}\,\cal M_{\odot}$ for the cluster.  These large uncertainties on the mass are a consequence of the small mass-sampling region for the IMF slope.  As a result the total mass that we present here should be treated with caution until deeper photometry can be analyzed.

\begin{figure}[h]
\includegraphics[width=\columnwidth]{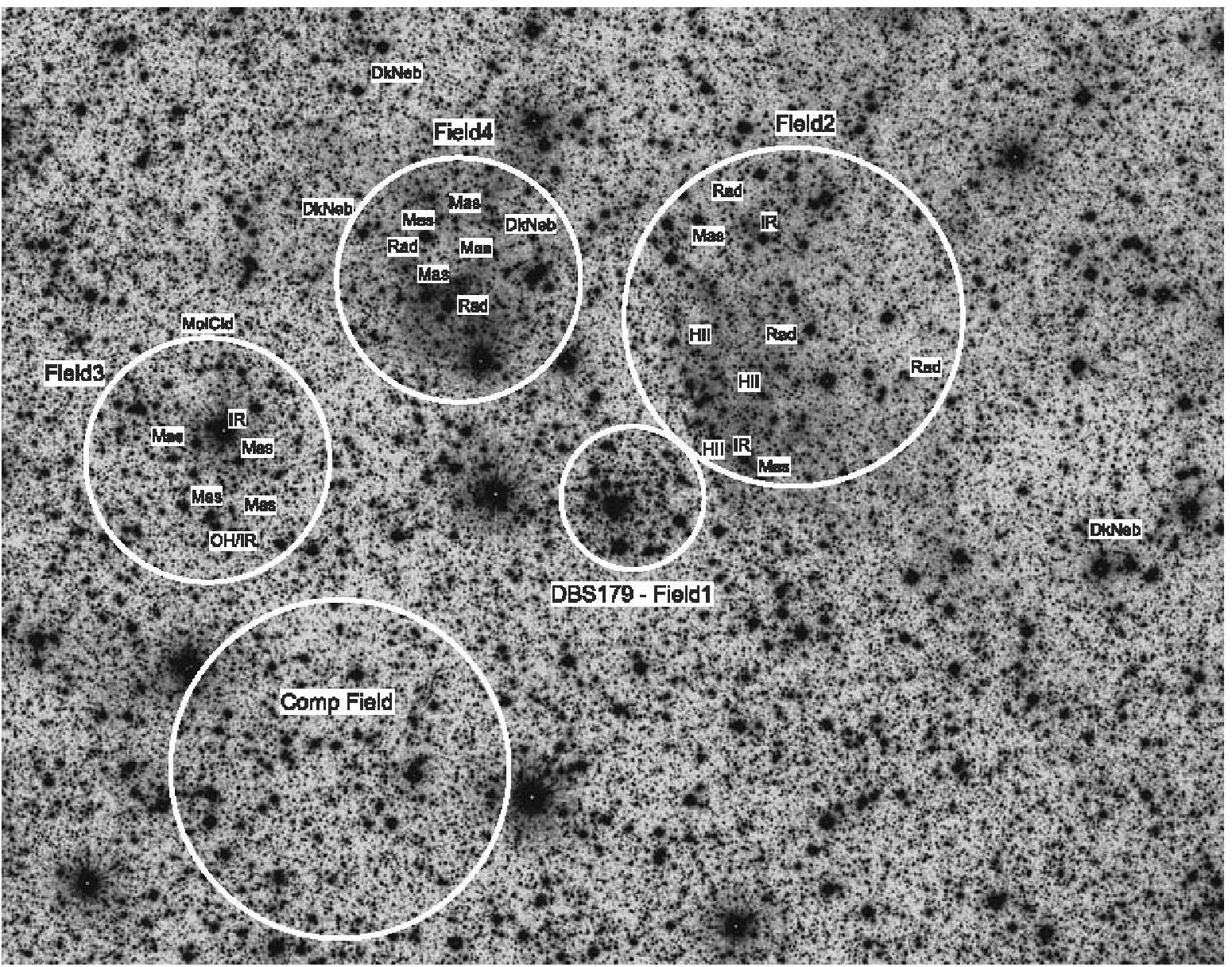}
\includegraphics[width=\columnwidth]{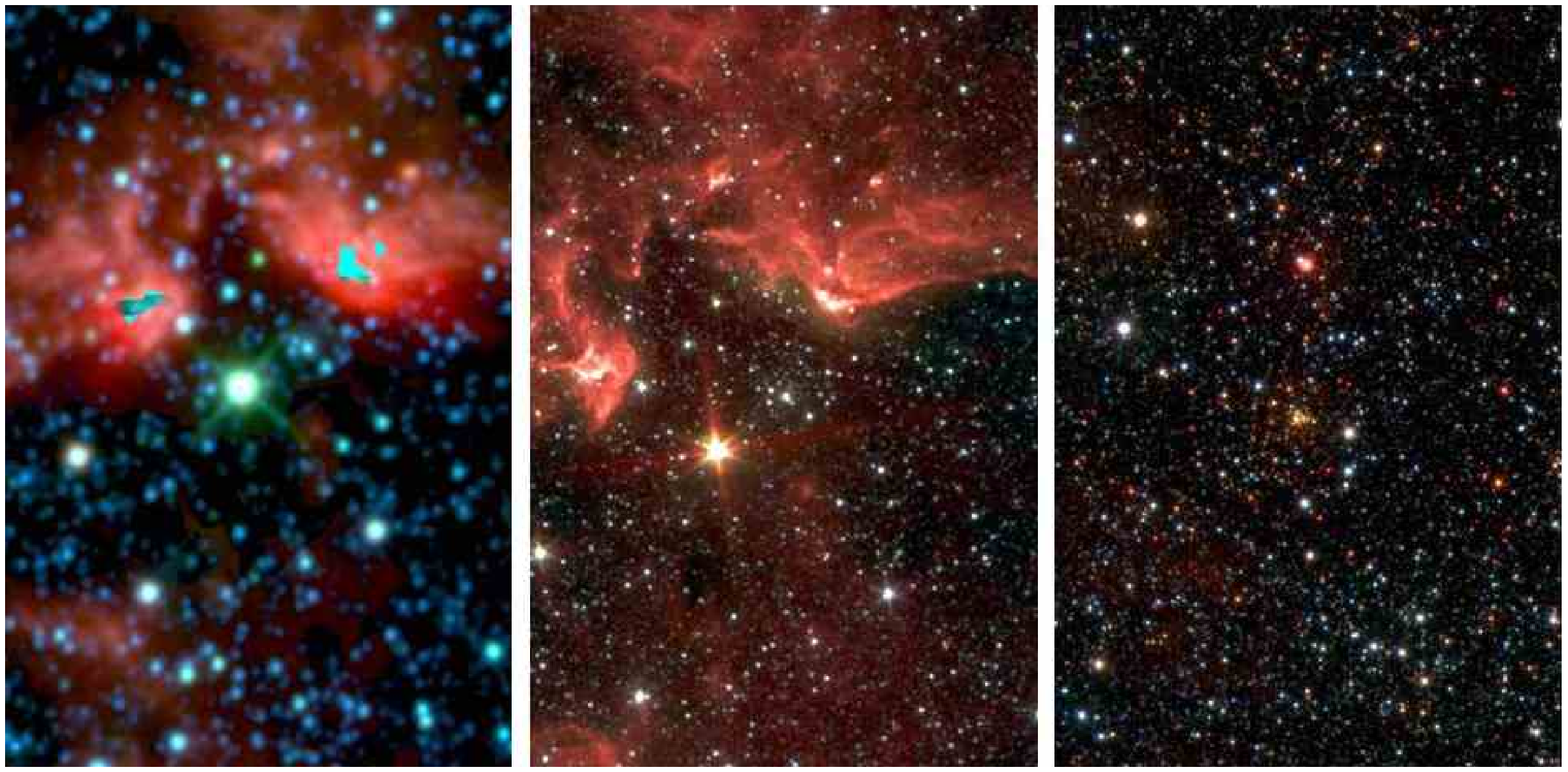}
\caption{Top panel: The VVV $K_{S}$ image of 15$\times$10\,arcmin region around DBS2003\,179 with known SIMBAD objects overplotted. Bottom panel: The WISE, GLIMPSE, and VVV true color images of 6$\times$10\,arcmin region.
}
\label{db179_all_regions}
\end{figure}

Finally, we show objects associated with the cluster star-forming complex. According to SIMBAD, within a field of radius=10\,arcmin around DBS2003\,179, we can see: three HII regions, IRAS 17079-3905, GAL 347.60+00.21, and GAL 347.6+00.2; four infrared sources, 2MASS J17113303-3910400, MSX5C G347.5845+00.2123, MSX5C G347.6316+00.2131, and IRAS 17078-3910; five radio sources, GRS 347.60 +00.20, GPSR 347.616+0.152, GPSR 347.602+0.246, GPA 347.63+0.20, and GPSR 347.629+0.149; ten masers; and three dark nebulae, which are indicators of ongoing star formation. They are overplotted on a VVV $K_{\rm S}$ band 15$\times$10\,arcmin image (Minniti al et. 2010, Saito et al. 2012) in Fig.~\ref{db179_all_regions}. These sources are located on the periphery of the cavity surrounding DBS2003\,179, clearly visible on GLIMPSE and WISE images.

\begin{figure}
\resizebox{\hsize}{!}{\includegraphics{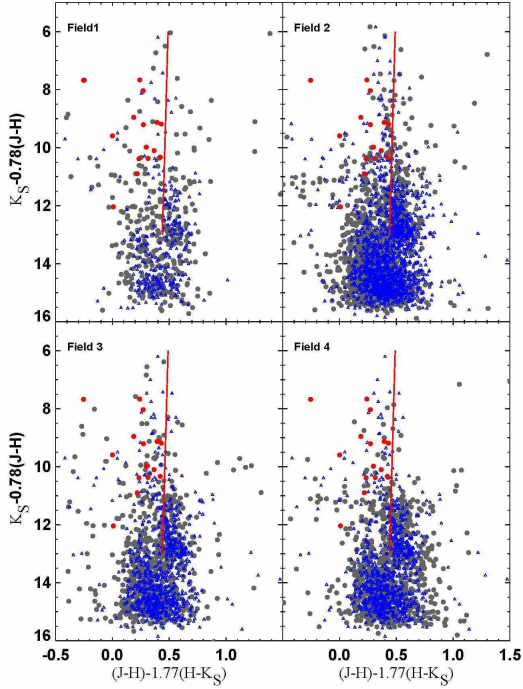}}\hspace{1cm}
\vspace{0.5cm}
\caption[]{The reddening-free color of $(J-$$H)$$-1.77(H-$$K_{\rm S})$ vs. $K_{\rm S}-$$0.78(J-$$H)$ magnitude.  The stars with spectra are marked with red filled dots, gray dots are all stars in the selected regions, the blue triangles represent the stars from the comparison filed. The solid line is a separation line between early OB stars and late-type main sequence stars and red clump giants (see the text). 
}
\label{db179cmd_regions_red_free}
\end{figure}

\begin{figure}
\begin{center}
\resizebox{9cm}{!}{\includegraphics{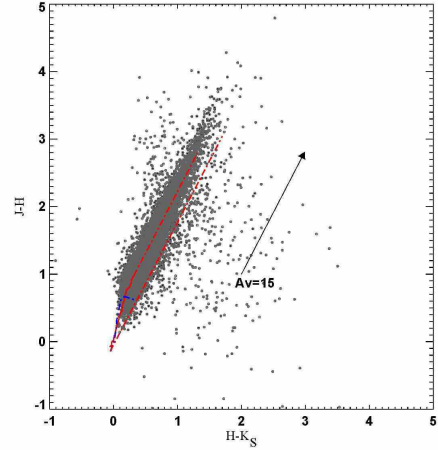}}
\caption[]{VVV observed $(J-$$H)$ vs. $(H-$$K_{\rm S})$ color-color diagram. Only stars with photometric errors $<$0.05 are plotted. The lines represent the sequence of zero-reddening stars of luminosity class\,{\sc{I}} (Koornneef 1983) and {\sc{V}} (Schmidt-Kaler 1982). The reddening vector for $A_{V}=15$\,mag is overplotted, and the dotted lines are parallel to the standard reddening vector.
}
\label{db179_cmd_vvv}
\end{center}
\end{figure}

Although not cluster members, the Objs. 7, 9 (discussed in Paper II), 18, and 19 are OB stars, indicating a diffuse population of early main sequence stars around the cluster. To search for additional objects of the same type we used the Negueruela et al. (2011) approach. The reddening-free parameter $Q=$$(J-$$H)$$-$1.77$(H-$$K_{S})$ was used to separate the early-type stars from late-type main sequence stars and red clump giants, in four regions associated with the cluster itself, and the star formation indicators (HII regions, infrared sources, masers, and radio sources). Figure~\ref{db179cmd_regions_red_free} shows this reddening free parameter vs. a corrected $K_{\rm S}$ magnitude for such selected areas, as well as a comparison field. Originally, Negueruela et al. (2011) defined $Q\leq$0.0$-$0.1 as a separating value for early-type stars. In our case, we defined this value using the right partition of the spectroscopically confirmed OB stars, namely $Q\leq$0.4$-$0.5. As can be seen in Fig.~\ref{db179cmd_regions_red_free}, there is no overdensity of these stars in Fields\,2, 3, and 4 with respect to the comparison field. In contrast, Field\,1 centered on DBS2003\,179 (with radius 1 arcmin) shows an overdensity. The spatial resolution of VVV images did not allow us to measure most of the DBS2003\, 179 cluster members due to crowding. Thus, there may be a larger OB association around the cluster, most likely preformed at the same time and from the same molecular cloud as DBS2003\,179. Interestingly, except for Obj\,6, only two other YSO objects have been reported to date in the close vicinity (15 arcmin radius) of the cluster. Analyzing the color-color diagram of the VVV frame, we can see some stars with infrared excess and/or PMS stars  (Fig.~\ref{db179_cmd_vvv}). However, our photometry is not deep enough to investigate this in more detail. 

In summary, we can conclude that, according to VVV images, there are no other younger and brighter than $K_{\rm S}$=18 mag star cluster and/or stellar groups in the close vicinity of DBS2003\,179. However, taking the total luminosity of modeled O and WR stars (approx 1.6\,10$^7$ ${L/L_\odot}$) into account, we should expect some continuous or burst protostar formation that is most probably triggered by DBS2003\,179 and the surrounding OB association. Indeed, several dust emission regions are visible on WISE and GLIMPSE. They will be analyzed in the next paper in our series.

\section{Summary}

We report new results for the massive evolved and main sequence members of the young galactic cluster DBS2003\,179.  
The cluster contains three late WN or WR/LBV stars (Obj\,4, Obj\,15, and Obj\,20 (MDM\,32)) and at least five OIf and five OV stars. According to the Hertzsprung-Russell diagram for DBS2003\,179 where the modeled parameters of all spectroscopic targets are plotted, the WR stars show masses above 85\,$\cal M_\odot$, the OIf stars are between 40\,$\cal M_\odot$ and 80\,$\cal M_\odot$, and the  main sequence O stars are $>$20\,$\cal M_\odot$. There are indications of binarity for Obj\,4, Obj\,11, and Obj\,20; Obj\,3 shows a variable spectrum; and Obj\,15 could be a star in transition. Using the newly added spectroscopic targets and photometry, the distance and mass of the cluster was redetermined, giving 7.9$\pm0.8$\,kpc and 2.5\,$10^4$\,$\cal M_\odot$, respectively. The WR stars fit 3 Myr isochrone well, while most of the O stars are placed around the 5 Myr isochrone. The slope of the IMF is consistent with the Salpeter slope. The cluster is surrounded by a continuous protostar formation region that was most probably triggered by DBS2003\,179. There is also an indication of diffuse OB association and pre-main sequence stars in the close vicinity of the cluster.

\begin{acknowledgements}
We are grateful to the anonymous referee, whose detailed comments and suggestions significantly improved the paper.
JB,JRAC, and FP are supported by the Chilean Ministry for the Economy, Development, and Tourism's Programa Iniciativa Cient\'{i}fica Milenio through grant P07-021-F,  awarded to The Milky Way Millennium Nucleus. Support for JB is provided by FONDECYT Reg. No. 1120601. JRAC is supported by GEMINI-CONICYT FUND No.32090002. FP is supported by GEMINI-CONICYT FUND No.32100020 and FONDECYT Reg. No. 1120601. RK is supported by the Centro de Astrofí\'isica de Valpara\'iso and Proyecto DIUV23/2009. The data used in this paper were obtained with ISAAC/VLT at the ESO Paranal Observatory. This publication makes use of data products from the Two Micron All Sky Survey, which is a joint project of the University of Massachusetts and the Infrared Processing and Analysis Center/California Institute of Technology, funded by the National Aeronautics and Space Administration and the National Science Foundation. This research made use of the SIMBAD database, operated at the CDS, Strasbourg, France.  MMH is grateful for the support provided to her by the ESO Visiting Scientist Program during an extended stay in late 2008 when the VLT spectra were reduced and acknowledges support by the National Science Foundation under contracts No. 0607497 and No. 1009550 to the University of Cincinnati. LNG is supported by CONACyT \#83016 and \#141530. We gratefully acknowledge use of data from the ESO Public Survey program ID 179.B-2002 taken with the VISTA telescope,  and data products from the Cambridge Astronomical Survey Unit.
\end{acknowledgements}

\end{document}